\documentclass[twocolumn,prl,superscriptaddress,showpacs,preprintnumbers,amsmath,amssymb]{revtex4-1}
\usepackage{graphicx}
\usepackage{color}

\begin{document}

\title{Kibble-Zurek mechanism in colloidal monolayers}

\author{Sven Deutschl\"ander}
\affiliation{University of Konstanz, D-78457 Konstanz, Germany}
\author{Patrick Dillmann}
\affiliation{University of Konstanz, D-78457 Konstanz, Germany}
\author{Georg Maret}
\affiliation{University of Konstanz, D-78457 Konstanz, Germany}
\author{Peter Keim}
\email{peter.keim@uni-konstanz.de}
\affiliation{University of Konstanz, D-78457 Konstanz, Germany}

\date{\today}

\begin{abstract}
The Kibble-Zurek mechanism describes the evolution of topological defect structures like domain walls, strings, and monopoles when a system is driven through a second order phase transition. The model is used on very different scales like the Higgs field in the early universe or quantum fluids in condensed matter systems. A defect structure naturally arises during cooling if separated regions are too far apart to `communicate' (e.g. about their orientation or phase) due to finite signal velocity. This results in separated domains with different (degenerated) locally broken symmetry. Within this picture we investigate the non-equilibrium dynamics in a condensed matter analogue, a two-dimensional ensemble of colloidal particles. In equilibrium it obeys the so called Kosterlitz-Thouless-Halperin-Nelson-Young (KTHNY) melting scenario with continuous (second-order like) phase transitions. The ensemble is exposed to a set of finite cooling rates covering roughly three orders of magnitude. Along this process, we analyze the defect and domain structure quantitatively via video microscopy and determine the scaling of the corresponding length scales as a function of the cooling rate. We indeed observe the scaling predicted by the Kibble-Zurek mechanism for the KTHNY universality class.
\end{abstract}


\maketitle

In the formalism of gauge theory with spontaneously broken symmetry, T.W.B. Kibble and colleagues postulated a cosmological phase transition during the cooling down of the early universe. This transition leads to degenerated states of vacua below a critical temperature, separated or dispersed by defect structures as domain walls, strings or monopoles~\cite{Zeldovich1975,Kibble1976,Kibble1980}. In the course of the transition, the vacuum can be described via an $N$-component, scalar order parameter $\phi$ (known as the Higgs field) underlying an effective potential
\begin{equation}
V=a\phi^2+b\left(\phi^2-\eta_0^2\right)^2
\label{eqnpot}
\end{equation}
where $a$ is temperature dependent, $b$ is a constant and $\eta_0$ is the modulus of $\left\langle\phi\right\rangle$ at $T=0$. For high temperatures, $V$ has a single minimum at $\phi=0$ (high symmetry) but develops a minimum `landscape' of degenerated vacua below a critical temperature $T_c$ (e.g. the so called sombrero shape for $N=2$). Cooling down from the high symmetry phase, the system undergoes a phase transition at $T_c$ into an ordered (low symmetry) phase with non-zero $\left\langle\phi\right\rangle$. For $T<T_c$ it holds
\begin{equation}
\left\langle\phi\right\rangle^2=\eta_0^2\left(1-T^2/T_c^2\right)=\eta^2(T)\ .
\label{eqnphi}
\end{equation}
Caused by thermal fluctuations, one can expect that below $T_c$, $\left\langle\phi\right\rangle$ takes different non-zero values in regions which are not connected by causality. The question now arising concerns the determination of the typical length scale $\xi_d$ of these regions and their separation. For a finite cooling rate, $\xi_d$ is limited by the speed of propagating information which is given by the finite speed of light defining an ultimate event horizon. Independent of the nature of the limiting causality, Kibble argued that as long as the difference in free energy $\Delta F$ (of a certain system volume) between its high symmetry state $\left\langle\phi\right\rangle=0$ and a possible finite value of $\left\langle\phi\right\rangle$ just below $T_c$ is less than $k_BT$, the volume can jump between both phases. The temperature at which $\Delta F=k_BT$ is called the Ginzburg temperature $T_G$ and the length scale $\xi_d$ of the initial (proto)domains is supposed to be equal to the correlation length at that temperature: $\xi_d=\xi(T_G)$~\cite{Kibble1976}.

\begin{figure}[b]
\centerline{\includegraphics[width=.48\textwidth]{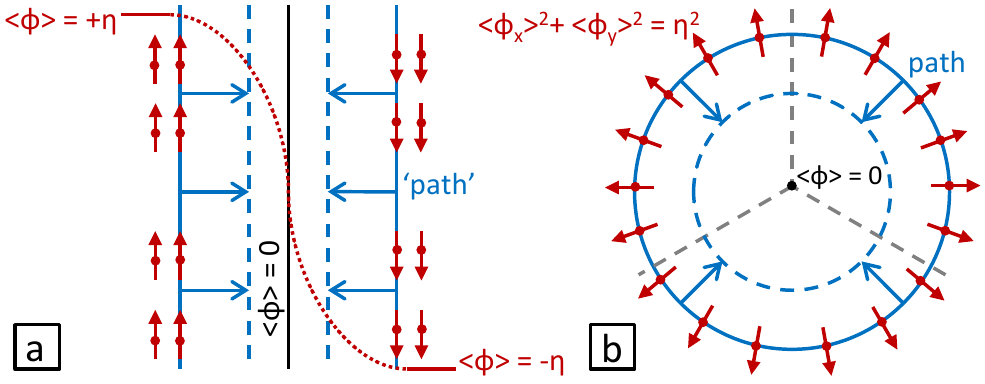}}
\caption{\label{phi}Emergence of defects in the Higgs-field which is illustrated with red vectors (shown in 2D for simplicity). a: For $N=1$ and $D=3$, domain walls can appear (strings for $D=2$). b: For $N=2$ and $D=3$, nontrivial topologies are strings (monopoles for  $D=2$). The defects are regions where the order parameter $\phi$ retains the high symmetry phase ($\left\langle\phi\right\rangle=0$) to 'moderate' between different degenerated orientations of the symmetry broken field.}
\end{figure}

The geometry of the defect network that separates the uncorrelated domains is given by the topology of the manifold of degenerated states that can exist in the low symmetry phase. Thus, it depends strongly on the dimensionality of the system $D$ and on the dimension $N$ of the order parameter itself. Regarding the square root of Eqn.~\ref{eqnphi}, the expectation value of a one-component order parameter ($N=1$) can only take two different low symmetry values $\left\langle\phi\right\rangle=\pm\eta(T)$ (like e.g. the magnetization in a 2D or 3D Ising model): The manifold of the possible states is `disconnected'. This has a crucial effect if one considers a mesh of symmetry broken domains where $\left\langle\phi\right\rangle$ is chosen randomly either $+\eta$ or $-\eta$. If two neighboring (but uncorrelated) domains have the same expectation value of $\left\langle\phi\right\rangle$, they can merge. Contrarily, domains with an opposite expectation value will be separated by a domain wall in 3D (or a `domain line' in 2D). At its center, the domain wall attains a value of $\left\langle\phi\right\rangle=0$, providing a continuous crossover of the expectation value between the domains (Fig.~\ref{phi}a). Consider now $N=2$: $\left\langle\phi\right\rangle$ can take any value on a circle, e.g. $\left\langle\phi\right\rangle^2=\left\langle\phi_x\right\rangle^2+\left\langle\phi_y\right\rangle^2=\eta^2(T)$ (all the order parameter values which are lying on the minimum-circle of the `sombrero' are degenerated). Since the manifold of possible low symmetry states is now connected, $\left\langle\phi\right\rangle$ can vary smoothly along a path (Fig.~\ref{phi}b). In a network of symmetry broken domains in two dimensions at least three domains (in Fig.\ref{phi}b separated by dashed lines) meet at a mutual edge. On a closed path around the edge, the expectation value $\left\langle\phi\right\rangle$ might be either constant along the path (for a global, uniform $\phi$) but can also vary by a multiple of $2\pi$ (in analogy to the winding numbers in liquid crystals). In the first case, the closed path can be reduced to a point with $\left\langle\phi\right\rangle \neq 0$ and no defect is built. If the path is shrunk in the second case, the field eventually has to attain $\left\langle\phi\right\rangle=0$ within the path and one remains with a monopole for $D=2$ or a string for $D=3$~\cite{Kibble1976,Kibble1980}. A condensed matter analogue would be a vortex of normal fluid with quantized circulation in superfluid helium. For $N=3$ and $D=3$, four domains can meet at a mutual \textit{point} and the degenerated solutions of the low temperature phase lie on a sphere: $\left\langle\phi\right\rangle^2=\left\langle\phi_x\right\rangle^2+\left\langle\phi_y\right\rangle^2+\left\langle\phi_z\right\rangle^2$. If the field now again varies circularly on a `spherical path' (all field arrows point radially outwards), a shrinking of this sphere leads to a monopole in three dimensions~\cite{Kibble1976,Kibble1980}.

W.H. Zurek extended Kibble's predictions and transferred his considerations to quantum condensed matter systems. He suggested that $^4$He should intrinsically develop a defect structure when quenched from the normal to the superfluid phase~\cite{Zurek1985,Zurek1993}. For superfluid $^4$He, the order parameter $\psi=|\psi|\exp(i\Theta)$ is complex with two independent components: magnitude $|\psi|$ and phase $\Theta$ (the superfluid density is given by $|\psi|^2$). A nontrivial, static solution of the equation of state with a Ginzburg-Landau potential yields $\psi=\psi_0(r)\exp(in\varphi)$ where $r$ and $\varphi$ are cylindrical coordinates, $n\in\mathbb{Z}$ and $\psi_0(0)=0$. This solution is called a vortex line, topologically equivalent to a string for the case $N=2$ we have discussed before. In the vicinity of the critical temperature during a quench from the normal fluid to the superfluid state, $\psi$ will be chosen randomly in uncorrelated regions leading to a string network of normal fluid vortices. In condensed matter systems, the role of the limiting speed of light is taken by the sound velocity (in $^4$He, the second sound). This leads to a finite speed of the propagation of order parameter fluctuations and sets a `sonic horizon'.

Zurek argued that the correlation length is `frozen-out' close to the transition point or even far before depending on the cooling rate~\cite{Zurek1985,Zurek1993}. Consider the divergence of the correlation length $\xi$ for a second-order transition, e.g. $\xi=\xi_0\left|\epsilon\right|^{-\nu}$ where $\epsilon=(T-T_c)/T_c$ is the reduced temperature. If the cooling is infinitely slow, the system behaves as in equilibrium: $\xi$ will diverge close to the transition and the system is a mono-domain. For an instantaneous quench, the system has minimal time to adapt to its surrounding: $\xi$ will be frozen-out at the beginning of the quench. For second order phase transitions, the divergence of correlation lengths is accompanied by the divergence of the correlation time $\tau=\tau_0\left|\epsilon\right|^{-\mu}$ which is due to the critical slowing down of order parameter fluctuations. If the time $t$ it takes to reach $T_c$ for a given cooling rate is larger than the correlation time, the system stays in equilibrium and the dynamic is adiabatic. Nonetheless, for every finite but nonzero cooling rate, $t$ eventually becomes smaller than $\tau$ and the system falls out of equilibrium before $T_c$ is reached. This is the so called freeze-out time $\hat{t}$, given when the correlation time equals the time it takes to reach $T_c$:
\begin{equation}
\hat{t}=\tau(\hat{t})\ .
\label{eqntau}
\end{equation}
The frozen out correlation length $\hat{\xi}$ is then set at the temperature $\hat{\epsilon}$ of the corresponding freeze-out time: $\hat{\xi} = \xi(\hat{\epsilon}) = \xi(\hat{t})$. For a linear temperature quench
\begin{equation}
\epsilon=(T-T_c)/T_c=t/\tau_q\
\label{eqntauq}
\end{equation}
with the quench time scale $\tau_q$, one observes $\hat{t}=\left(\tau_0\tau_q^{\mu}\right)^{1/(1+\mu)}$ and
\begin{equation}
\hat{\xi}=\xi(\hat{t})=\xi_0\left(\tau_q/\tau_0\right)^{\nu/(1+\mu)}\ .
\label{eqnpl}
\end{equation}
For the GL model ($\nu=1/2$, $\mu=1$) one finds the scaling $\hat{\xi}\sim\tau_q^{1/4}$ while a renormalization group correction ($\nu=2/3$) leads to $\hat{\xi}\sim\tau_q^{1/3}$~\cite{Zurek1985,Zurek1993}.

A frequently used approximation is that when the adiabatic regime ends at $\hat{t}$ \textit{before} the transition, the correlation length can not follow the critical behavior until $\tau$ again exceeds the time $t$ when $T_c$ is passed. Given a symmetric divergence of $\tau$ around $T_c$, this is the time $\hat{t}$ \textit{after} the transition. The period in between is known as the impulse regime in which the correlation length is assumed not to evolve further. A recent analytical investigation, however, suggests that in this period, the system falls into a regime of critical coarse graining~\cite{Biroli2010}. There, the typical length scale of correlated domains continues to grow since local fluctuations are still allowed and the system is out of equilibrium. On the other side, numerical studies in which dissipative contributions and cooling rates were alternatively varied before and after the transition indicate that the final length scale of the defect and domain network is entirely determined after the transition~\cite{Antunes2006}. Several efforts have been made to provide experimental verification of the Kibble-Zurek mechanism in a variety of systems, e.g. in liquid crystals~\cite{Chuang1991} (the transition is weakly first order but the defect network can easily observed with cross polarization microscopy), superfluid $^3$He~\cite{Baeuerle1996}, superconducting systems~\cite{Carmi2000}, convective, intrinsically out of equilibrium systems~\cite{Miranda2013}, multiferroics~\cite{Chae2012}, quantum systems~\cite{Xu2014}, ion crystals~\cite{Ulm2013,Pyka2013}, and Bose-Einstein condensates~\cite{Lamporesi2013} (the latter two systems contain the effect of inhomogeneities due to e.g. temperature gradients). A detailed review concerning the significance and limitations of these experiments can be found in~\cite{Campo2014}.

In this experimental study, we test the validity and applicability of the Kibble-Zurek mechanism in a two-dimensional colloidal model system whose equilibrium thermodynamics follow the microscopically motivated Kosterlitz-Thouless-Halperin-Nelson-Young (KTHNY) theory. This theory predicts a continuous, two-step melting behavior whose dynamics, however, are quantitatively different from phenomenological second-order phase transitions described by the Ginzburg-Landau model. We applied cooling rates over roughly three orders of magnitude for which we changed the control parameter with high resolution and homogeneously throughout the sample without temperature gradients. Single particle resolution provides a quantitative determination of defect and domain structures during the entire quench procedure, and the precise knowledge of the equilibrium dynamics allows to determine the scaling behavior of corresponding length scales at the freeze-out times. In the following, we validate that the Kibble-Zurek mechanism can be successfully applied to the KTHNY universality class.
\begin{figure}[b]
\centerline{\includegraphics[width=.48\textwidth]{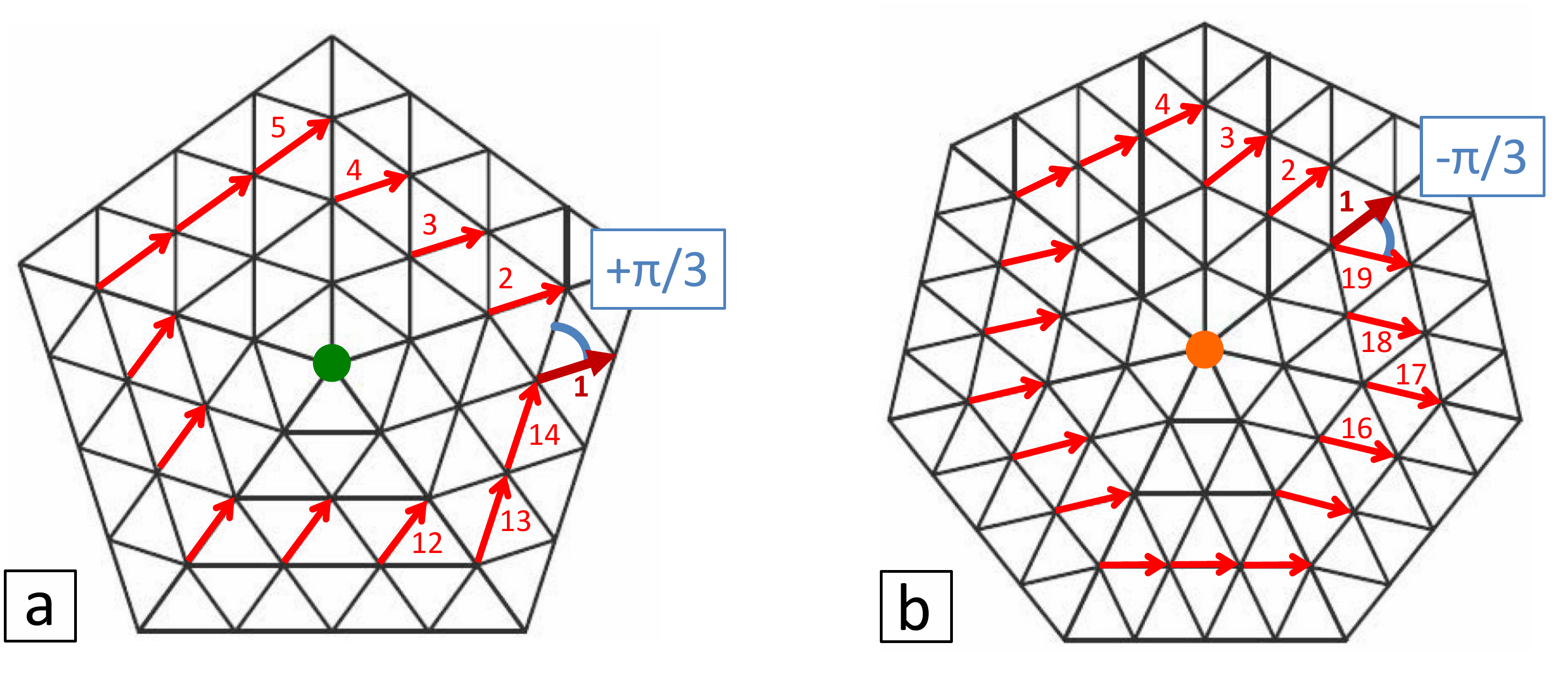}}
\caption{\label{disclination} Sketch of a fivefold oriented (a) and sevenfold oriented (b) disclination. The red arrows illustrate the change in bond angle (blue) when circling on an anti clock-wise path around the defect.}
\end{figure}

\section{Defects and symmetry breaking in 2D crystallization}

The closed packed crystalline structure in two dimensions is a hexagonal crystal with sixfold symmetry. The thermodynamics of such a crystal can analytically be described via the KTHNY theory, a microscopic, two-step melting scenario (including two continuous transitions) which is based on elasticity theory and a renormalization group analysis of topological defects~\cite{Kosterlitz1973,Young1979,Nelson1979}. In the KTHNY formalism, the orientationally long-range-ordered crystalline phase melts at a temperature $T_m$ via the dissociation of pairs of dislocations into an hexatic fluid, which is unknown in 3D systems. This fluid is characterized by quasi-long-range orientational but short-range translational order. In a triangular lattice, dislocations are point defects and consist of two neighboring particles with five and seven nearest neighbors, respectively, surrounded by sixfold coordinated particles. At a higher temperature $T_i$, dislocations start to unbind further into isolated disclinations (a disclination is a particle with five \textit{or} seven nearest neighbors surrounded by sixfold coordinated particles) and the system enters an isotropic fluid with short-range orientational and translational order. A suitable orientational order parameter is the local bond order field $\psi_{6}(\vec{r}_j,t) = n_j^{-1}\sum_{k}e^{i6\theta_{jk}(t)}=|\psi_{6}(\vec{r}_j,t)|e^{i\Theta_j(t)}$ which is a complex number with magnitude $|\psi_{6}(\vec{r}_j,t)|$ and phase $\Theta_j(t)$ defined at the discrete particle positions $\vec{r}_j$. $\Theta_j(t)$ is the average bond orientation for a specific particle. The $k$-sum runs over all $n_j$ nearest neighbors of particle $j$, and $\theta_{jk}$ is the angle of the $k$th bond with respect to a certain reference axis. If particle $j$ is perfectly sixfold coordinated (e.g. all $\theta_{jk}(t)$ equal an ascending multiple of $\pi/3$), the local bond order parameter attains $|\psi_{6}(\vec{r}_j,t)|=1$. A five- or sevenfold coordinated particle yields $|\psi_{6}(\vec{r}_j,t)|\gtrsim 0$. The three different phases can be distinguished via the spatial correlation $g_6(r)=\left\langle \psi_6^*(\vec{0})\psi_6(\vec{r_j})\right\rangle$ or temporal correlation $g_6(t)=\left\langle \psi_6^*(0)\psi_6(t)\right\rangle$ of the local bond order parameter. For large $r$ and $t$, $g_6(r,t)$ attains a finite value in the (mono)crystalline phase, decays algebraically in the hexatic fluid, and exponentially $\sim\exp(-r/\xi_6)$ and $\sim\exp(-t/\tau_6)$ in the isotropic fluid~\cite{Nelson1979,Nelson1983}. Unlike second-order phase transitions where correlations typically diverge algebraically, the orientational correlation length $\xi_6$ and time $\tau_6$ diverge in the KTHNY formalism exponentially at $T_i$:
\begin{equation}
\xi_6\sim\exp(a\left|\epsilon\right|^{-1/2})\ \ \ \ \textrm{and}\ \ \ \ \tau_6\sim\exp(b\left|\epsilon\right|^{-1/2})\ ,
\label{eqnxitau}
\end{equation}
where $\epsilon=(T-T_i)/T_i$, and $a$ and $b$ are constants~\cite{Nelson1979,Watanabe2004}. This peculiarity is the reason why KTHNY-melting is named continuous instead of second-order. In equilibrium, the KTHNY scenario has been verified successfully for our colloidal system in various experimental studies~\cite{Zahn1999,Keim2007,Deutschlander2014}.

\begin{figure}[t]
\centerline{\includegraphics[width=.5\textwidth]{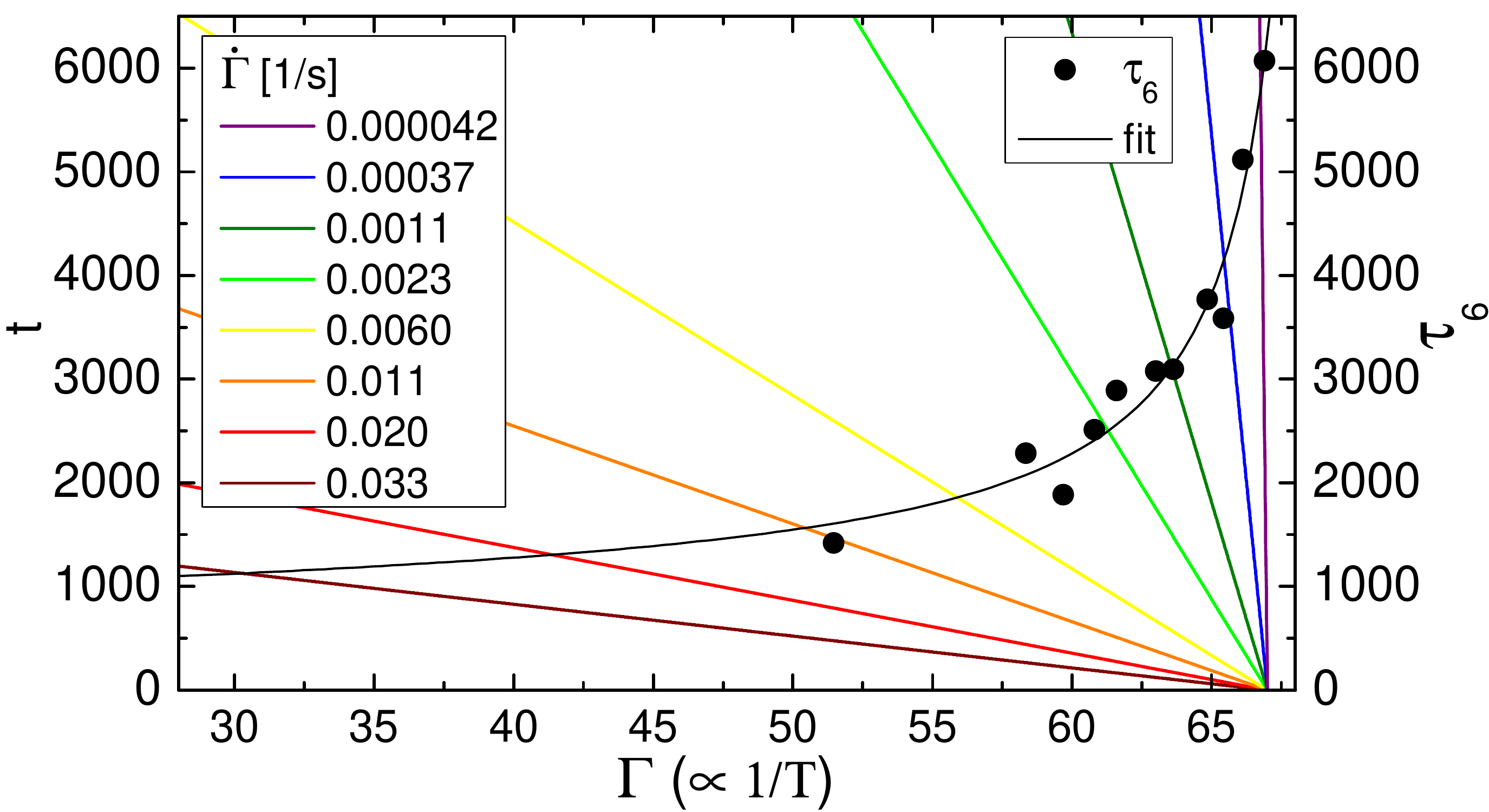}}
\caption{\label{time} Orientational correlation time $\tau_6$ (experimental data and fit according to Eqn.~\ref{eqnxitau}) and the time $t$ left until the transition temperature is reached for different cooling rates (colored straight lines, Eqn.~\ref{eqnt}) as function of inverse temperature $\Gamma$ (small $\Gamma$ correspond to large temperatures and vice versa). The intersections define the freeze-out interaction values $\hat{\Gamma}=\Gamma(\hat{t})$.}
\end{figure}

To transfer this structural 2D phase behavior into the framework of the Kibble-Zurek mechanism, we start in the high temperature phase (isotropic fluid) and describe the symmetry breaking with the spatial distribution of the bond order parameter. Since in 2D the local symmetry is sixfold in the crystal and the fluid, the isotropic phase is a mixture of sixfold and equally numbered five- and sevenfold particles (other coordination numbers are extremely rare and can be neglected). During cooling, isolated disclinations combine to dislocations which for `infinite' slow cooling rates can annihilate into sixfold particles with an uniform director field. This is given by a global phase, characterizing the orientation of the crystal axis. Spontaneous symmetry breaking implies that all possible global crystal orientations are degenerated and the Kibble-Zurek mechanism predicts that in the presence of critical fluctuations the system can not gain a global phase at finite cooling rates: Locally, symmetry broken domains will emerge which will have different orientations in causally separated regions. The final state is a polycrystalline network with frozen-in defects. As in the case of superfluid $^4$He, $\psi_6(\vec{r}_j,t)$ is complex with two independent components ($N=2$). Consequently, we expect to observe monopoles in two dimensions. The phase of $\psi_6(\vec{r}_j,t)$ is invariant under a change in the particular bond angles of $\Delta\theta_{jk}(t)=\pm n\pi/3$ ($n\in\mathbb{N}$) which is caused by the sixfold orientation of the triangular lattice. Similar to the Higgs field or the superfluid, one can not consider a closed (discrete) path in $\psi_6(\vec{r},t)$ on which $\theta_{jk}(t)$ changes by an amount of $\pm\pi/3$ leaving the orientational field invariant. Reducing this path to a point, $\psi_6(\vec{r},t)$ must tend towards zero at the center to maintain continuity. Since the orientational field is defined at discrete positions, the defect is a single particle marked as a monopole of the high symmetry phase. In fact, this coincides with the definition of disclinations in the KTHNY formalism~\cite{Nelson1979}: The particle at the center is an isolated five- or sevenfold coordinated site.
Fig.~\ref{disclination} illustrates this for a bond on a closed path. Going counter clock-wise around the defect, the bond angle changes by an amount of $+\pi/3$ for a fivefold (Fig.~\ref{disclination}a) and by $-\pi/3$ for a sevenfold site (Fig.~\ref{disclination}b). (In principle, also larger changes in $\theta_{jk}(t)$ are possible, e.g. for $n=2$, a four- or eightfold oriented site, but these are extremely rare.) In KTHNY theory the monopoles (disclinations) combine to dipoles (dislocations) which can only annihilate completely if their orientation is exactly antiparallel. At finite cooling rates they arrange in chains, separating symmetry broken domains of different orientation: Chains of dislocations can be regarded as strings or `2D domain walls'.

\section{Colloidal monolayer and cooling procedure}

Our colloidal model system consists of polystyrene beads with diameter $\sigma=4.5\;\mu\textrm{m}$, dispersed in water and sterically stabilized with the soap sodium dodecyl sulfate (SDS). The beads are doped with iron oxide nanoparticles which results in a superparamagnetic behavior and a mass density of 1.7 kg/dm$^3$. The colloidal suspension is sealed within a millimeter sized glass cell where sedimentation leads to the formation of a monolayer of beads on the bottom glass plate. The whole layer consists of $>10^5$ particles and in a $1158\times865\;\mu\textrm{m}^2$ sub window $\approx5700$ particles are tracked with a spatial resolution of sub-micrometers and a time resolution in the order of seconds. The system is kept at room temperature and exempt from density gradients due to a months-long precise control of the horizontal inclination down to $\mu$rad. The potential energy can be tuned by an external magnetic field $H$ applied perpendicular to the monolayer which induces a repulsive dipole-dipole interaction between the particles. The ratio between potential energy $E_{\textrm{mag}}$ and thermal energy $k_BT$,
\begin{equation}
\Gamma=\frac{E_{\textrm{mag}}}{k_BT}=\frac{\mu_0\left(\pi n\right)^{3/2}\left(\chi H\right)^2}{4\pi k_BT}\ ,
\label{eqn_gamma}
\end{equation}
acts as inverse temperature (or dimensional pressure for fixed volume and particle number). $n=1/a_0^2$ is the 2D particle density with a mean particle distance $a_0\approx13$ $\mu$m, and $\chi=1.9\cdot10^{-11}\;\textrm{Am}^2/\textrm{T}$ is the magnetic susceptibility of the beads. $\Gamma$ is the thermodynamic control parameter: A small magnetic field corresponds to a large temperature and vice versa. Measured values of the equilibrium melting temperatures are $\Gamma_m\approx70.3$ for the crystal/hexatic transition and $\Gamma_i\approx67.3$ for the hexatic/isotropic transition~\cite{Deutschlander2014}. The cooling procedure is the following: We equilibrate the system deep in the isotropic liquid at $\Gamma_{0}\approx25$ and apply linear cooling rates $\dot{\Gamma}=\Delta\Gamma/\Delta t$ deep into the crystalline phase up to $\Gamma_{\textrm{end}}\approx100$ thenceforward we let the system equilibrate. We perform different rates, ranging over almost three decades from $\dot{\Gamma}=0.000042\;1/\textrm{s}$ up to $\dot{\Gamma}=0.0326\;1/\textrm{s}$. The slowest cooling rate corresponds to a quench time of $\approx19$ days and the fastest one to $\approx40$ minutes. We would like to emphasize that with the given control parameter there is no heat transport from the surface as in 3D bulk material. The lack of gradients rules out a temperature gradient assisted annealing of defects which might be present in inhomogeneous systems.

\begin{figure}[b]
\centerline{\includegraphics[width=.48\textwidth]{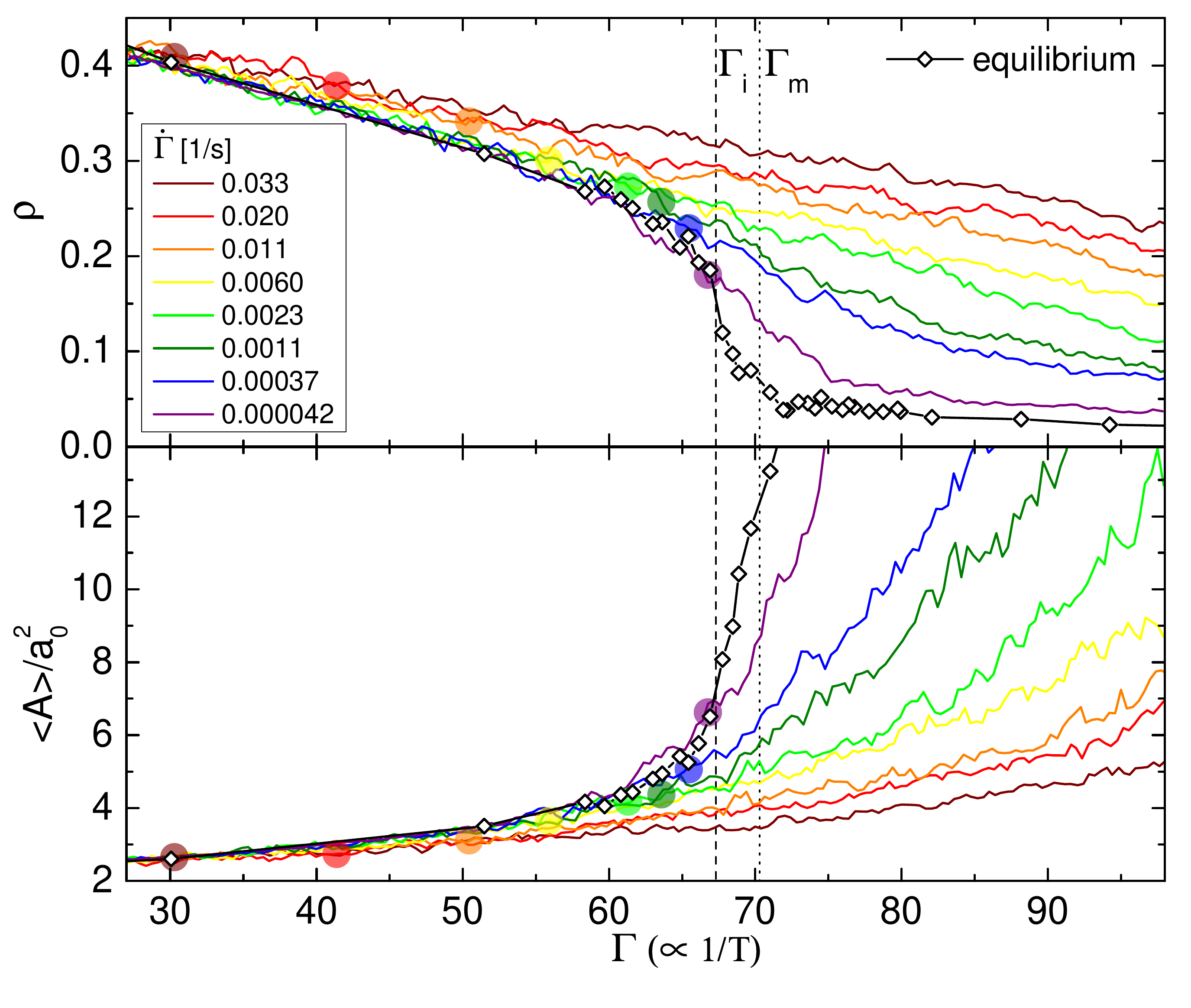}}
\caption{\label{defectsandcluster} Defect number density $\rho$ and average domain size $\left\langle A\right\rangle$ (in units of $a_0^2$) as a function of $\Gamma$ ($\propto1/T$) during cooling from small $\Gamma$ (= `hot' on the left side) to large $\Gamma$ (= `cold' on the right side). The curves cover the complete range of cooling rates from $\dot{\Gamma}=0.000042$ to $\dot{\Gamma}=0.033$ and are averaged within an interval $\Delta\Gamma=0.4$. Big dots mark the freeze-out temperatures $\hat{\Gamma}=\Gamma(\hat{t})$ (colored correspondingly to $\dot{\Gamma}$). Open symbols show the equilibrium melting behavior (lines are guide to the eye).}
\end{figure}

\section{Structure and dynamics of defects and domains}

The key element of the Kibble-Zurek mechanism is a frozen-out correlation length $\hat{\xi}$ as the system falls out of equilibrium at the freeze-out time $\hat{t}$. For slow cooling rates, the system can follow adiabatically closer to the transition (large $\hat{\xi}$) than for fast rates where the systems reaches the freeze-out time earlier (small $\hat{\xi}$). To find $\hat{t}$, we determine the orientational correlation time $\tau_6$ according the KTHNY theory by fitting $g_6(t)\sim\exp(-t/\tau_6)$ in the isotropic fluid for independent equilibrium measurements. The data for $\tau_6$ as well as a fit with
\begin{equation}
\tau_6=\tau_{0}\exp(b_{\tau}\left|1/\Gamma-1/\Gamma_c\right|^{-1/2})
\label{eqnfit}
\end{equation}
is shown on the right axis of Fig.~\ref{time}. The time left to the isotropic-hexatic transition is given by
\begin{equation}
t=(\Gamma_i-\Gamma)/\dot{\Gamma}
\label{eqnt}
\end{equation}
and is also plotted in Fig.~\ref{time} (left axis) for various cooling rates including the slowest and the fastest one. The points of intersection
\begin{equation}
\hat{t}=\tau_0\exp\left(b_{\tau}\left|1/\Gamma(\hat{t},\dot{\Gamma})-1/\Gamma_c\right|^{-1/2}\right)
\label{eqnt2}
\end{equation}
define the freeze-out `temperatures' $\hat{\Gamma}=\Gamma(\hat{t})$.

The length scale of the defect network can be measured by the concentration $\rho$ of defects (counting all not sixfold coordinated particles) in the $\psi_6(\vec{r},t)$-field. Fig.~\ref{defectsandcluster}~(upper plot) shows the evolution of $\rho$ for the same cooling rates $\dot{\Gamma}$ as in Fig.~\ref{time}, as well as for the equilibrium (melting) behavior~\cite{Deutschlander2014}. One recognizes that the course of $\rho$ deviates from the equilibrium case in advance of the isotropic/hexatic transition at $\Gamma_i\approx67.3$. This happens at different times for distinct cooling rates and marks the end of the adiabatic regime. Within the noise, deviations from the equilibrium behavior start at the temperature $\hat{\Gamma}$ given by the freeze-out time $\hat{t}$ (big colored dots). Beyond the adiabatic regime the defect density decreases which is an indication of critical coarse graining as predicted in~\cite{Biroli2010}. At $\Gamma_i$ (and also $\Gamma_m$) the slope of the curves increases with decreasing cooling rate indicating a further evolution but the system cannot perform critical fluctuations.

The domain structure, on the other hand, can be characterized quantitatively by analyzing symmetry broken domains with similar phase of $\psi_{6}(\vec{r}_j,t)$. According to~\cite{Dillmann2013}, we define a particle to be part of a symmetry broken domain if the following three conditions are fulfilled for the particle itself and at least one nearest neighbor: 1) The magnitude $|\psi_{6}(\vec{r}_j,t)|$ of the local bond order parameter must exceed $0.6$ for both neighboring particles, 2) the bond length deviation of neighboring particles is less than $10\%$ of the average particle distance $a_0$, and 3) the variation in the average bond orientation $\Delta\Theta_{ij}(t)=\left|Im[\psi_6(\vec{r}_i)]-Im[\psi_6(\vec{r}_j)]\right|$ of neighboring particles $i$ and $j$ must be less than $14^\circ$ (less than $14^\circ/6$ in real space). Simply connected domains of particles which fulfill all three criteria are merged to a local symmetry broken domain. If a particle does not satisfy these conditions in respect to a neighboring particle, it is assigned to the high symmetry phase (almost all defects are identified as such due to their small value of $|\psi_6(\vec{r}_j,t)|$). Fig.~\ref{defectsandcluster}~(lower plot) shows the evolution of the ensemble average domain size $\left\langle A\right\rangle$ as a function of $\Gamma$. We observe a behavior analogue to $\rho$: Domain formation significantly deviates from the equilibrium case before $\Gamma_i$, namely around the freeze-out temperature $\hat{\Gamma}$ of the corresponding cooling rates $\dot{\Gamma}$. To compare both networks in the following, we define the dimensionless lengths
$\xi_{\textrm{def}}=\rho^{-1/2}$ and $\xi_{\textrm{dom}}=(\left\langle A\right\rangle/a^2_0)^{1/2}$
which display the characteristic length scales in units of $a_0$.

\begin{figure}[t]
\centerline{\includegraphics[width=0.5\textwidth]{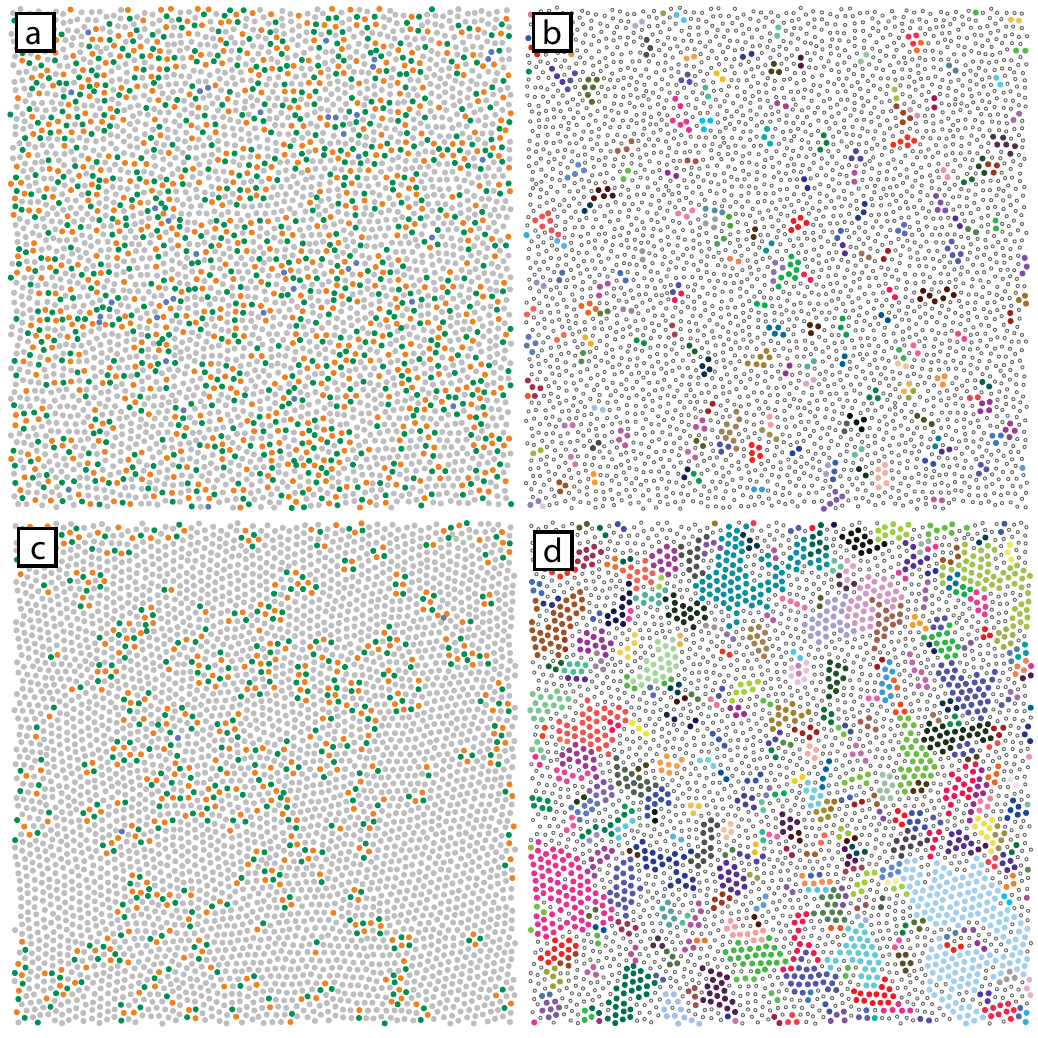}}
\caption{\label{figsnap1} Snapshot sections of the colloidal ensemble ($992\times960\;\mu\textrm{m}^2,\approx4000$ particles) illustrating the defect (a,c) and domain configurations (b,d) at the freeze out temperature $\hat{\Gamma}$ for the fastest (a,b: $\dot{\Gamma}=0.0326\;1/\textrm{s}$, $\hat{\Gamma}\approx30.3$) and slowest cooling rate (c,d: $\dot{\Gamma}=0.000042\;1/\textrm{s}$, $\hat{\Gamma}\approx66.8$). The defects are marked as follows: Particles with five nearest neighbors are colored red, seven nearest neighbors green and other defects blue. Sixfold coordinated particles are colored grey. Different symmetry broken domains are colored individually and high symmetry particles are displayed by smaller circles.}
\end{figure}

\begin{figure}[t]
\centerline{\includegraphics[width=0.5\textwidth]{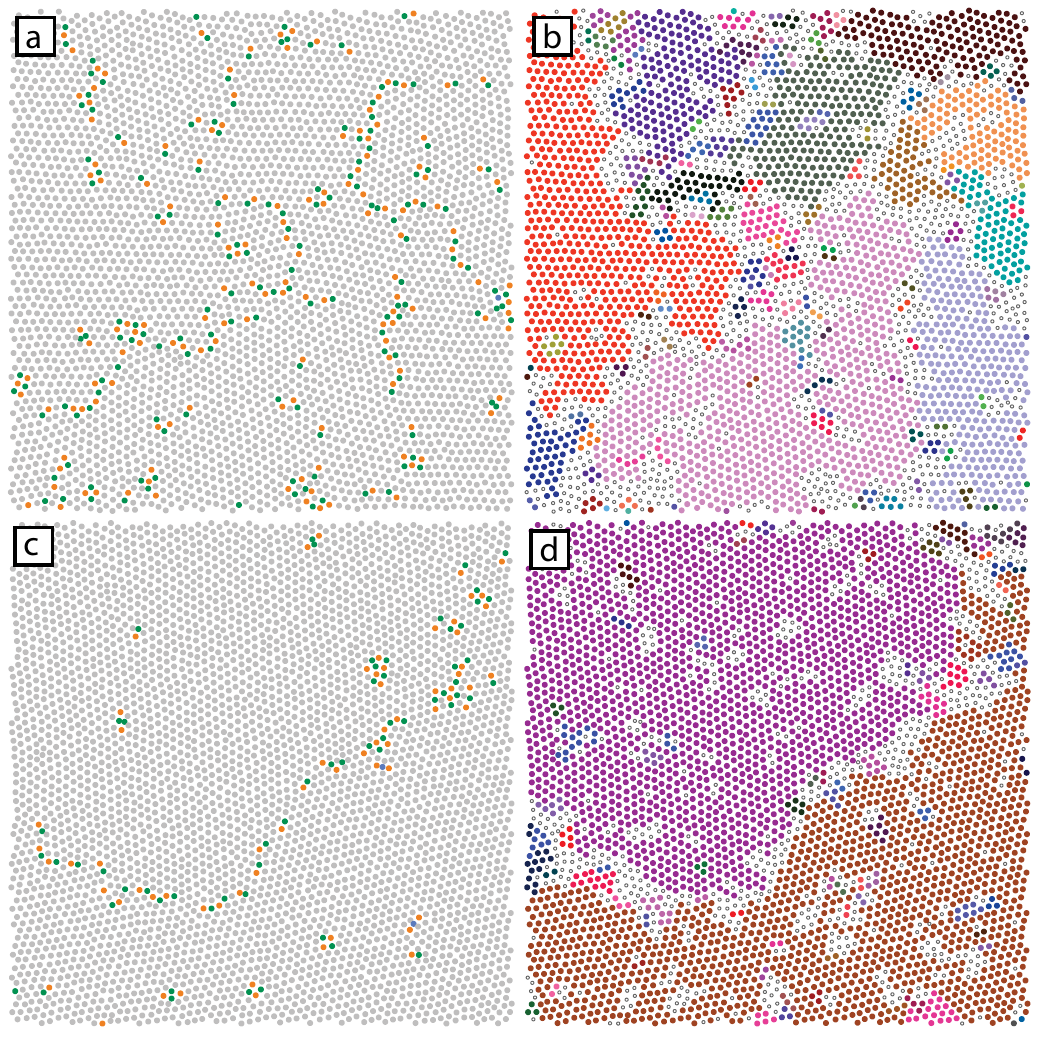}}
\caption{\label{figsnap2} Snapshot sections of the colloidal ensemble illustrating the defect (a,c) and domain configurations (b,d) after quasi-equilibration of the system for the fastest (a,b: $\dot{\Gamma}=0.0326\;1/\textrm{s}$, $\Gamma_{\textrm{end}}\approx105$) and slowest cooling rate (c,d: $\dot{\Gamma}=0.000042\;1/\textrm{s}$, $\Gamma_{\textrm{end}}\approx98$). The system size and the labeling of defects and domains is the same as in Fig.~\ref{figsnap1}.}
\end{figure}
Colloidal ensembles offer the unique possibility to monitor the system and its domain and defect structure on single particle level. Fig.~\ref{figsnap1} illustrates both (left row for defects, right row for domains) at the freeze-out temperature $\hat{\Gamma}$ for the fastest (a,b) and the slowest (c,d) cooling rate. For $\dot{\Gamma}=0.0326\;1/\textrm{s}$ (a,b) where $\hat{t}$ is already reached at $\hat{\Gamma}=30.3$, the defect density is large as is the number of high symmetry particles. However, there is a significant number of sixfold coordinated particles and a few orientationally ordered domains (to accord for finite size effects, we will exclude domains which hit the border of the field of view when evaluating $\xi_{\textrm{dom}}$ at $\hat{\Gamma}$). At this point the length scales are $\xi_{\textrm{def}}=1.56\pm0.01$ and $\xi_{\textrm{dom}}=1.56\pm0.03$. 
For the slowest cooling rate $\dot{\Gamma}=0.000042\;1/\textrm{s}$ (c,d) where $\hat{\Gamma}=66.7$, the mean distance between defects as well as the typical domain size is significantly larger compared to the fastest cooling rate. We observe $\xi_{\textrm{def}}=2.36\pm0.07$, and $\xi_{\textrm{dom}}=2.30\pm0.09$.

To allow relaxation of the defect and domain structure after the freeze-out time~\cite{Biroli2010}, we keep the temperature constant after $\Gamma_{\textrm{end}} \approx 100$ is reached. Fig.~\ref{figsnap2} shows the defect and domain configurations after an equilibration time of $\approx5$ hours for the fastest cooling rate (a,b) where the quench time was $\approx40$ minutes, and after an equilibration time of $\approx3$ days for the slowest rate (c,d) where the quench time was $\approx19$ days. The different length scale of the defect and symmetry broken domain network in respect to the cooling rate is clearly visible: While we observe a large number of domains for fast cooling, slow cooling results in merely two large domains separated by a single grain boundary. The final evolution will be given by classical coarse graining. The ground state is known to be a monodomain but its observation lies beyond experimental accessible times for our system.

\begin{figure}[b]
\centerline{\includegraphics[width=.48\textwidth]{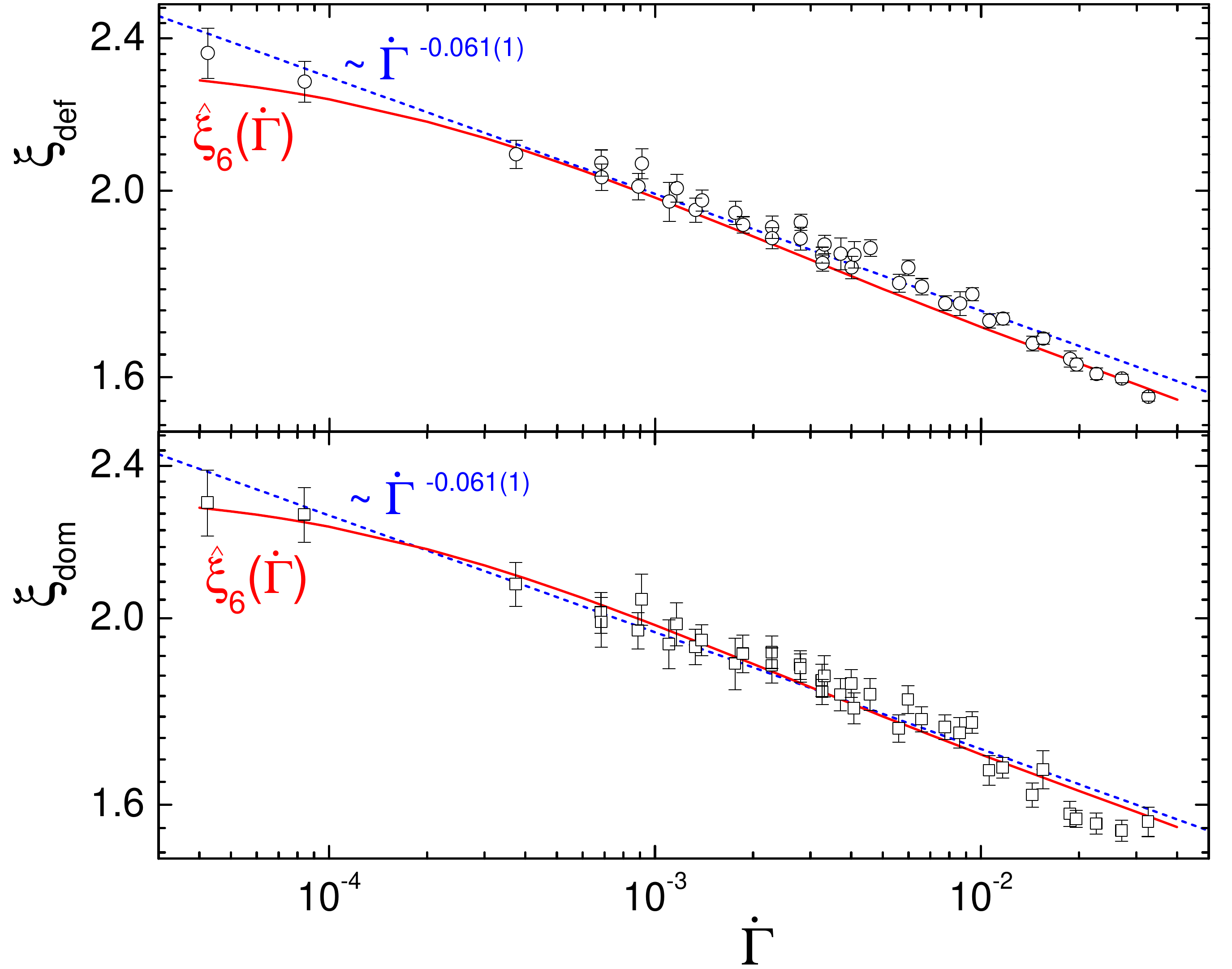}}
\caption{\label{quench} The length scale of the defect $\xi_{\textrm{def}}$ and domain network $\xi_{\textrm{dom}}$ is plotted as a function of the cooling rate $\dot{\Gamma}$ (open symbols). Red lines are numerical solutions of the transcendental equation following the freeze-out condition for the KTHNY-like divergences (see text for definition). For comparison, dashed blue lines are power law fits predicted by standard Kibble-Zurek mechanism which show the same algebraic exponent $\kappa\approx0.06$ for $\xi_{\textrm{def}}$ and $\xi_{\textrm{dom}}$.}
\end{figure}

\section{Scaling behavior}

The main prediction of the Kibble-Zurek mechanism is a power law dependence of the frozen-out correlation length $\hat{\xi}$ as function of $\tau_q$ (Eqn.~\ref{eqnpl}) which results from the algebraic divergence of the correlation, presuming Eqn.~\ref{eqntau} and Eqn.~\ref{eqntauq}. In KTHNY-melting, $\xi_6$ and $\tau_6$ diverge exponentially and one has to solve Eqn.~\ref{eqnt2} to find the implicit dependency $\hat{t}(\dot{\Gamma})$. We did this numerically for discrete values in the range $4\cdot10^{-5}\leq\dot{\Gamma}\leq4\cdot10^{-2}$ and determined the frozen-out orientational correlation length $\hat{\xi}_6$ for a scaling $\tau_6/\tau_B=c\:\xi_6^z$ with the dynamical exponent $z$ \cite{Watanabe2004}. Here, $\tau_B \approx 171.6\;\textrm{s}$ is the Brownian time which is the time a single particle needs to diffuse its own diameter. Using Eqn.~\ref{eqnfit} one finds with $\Gamma(t,\dot{\Gamma})$ from Eqn.~\ref{eqnt} the expression
\begin{equation}
\hat{\xi}_6(\dot{\Gamma})=\left(\frac{\tau_0}{c\tau_B}\right)^{1/z}\exp\left(\frac{b_{\tau}}{z}\left|\frac{\Gamma_c-\Gamma_i+\dot{\Gamma}\hat{t}(\dot{\Gamma})} {\Gamma_c\Gamma_i-\Gamma_c\dot{\Gamma}\hat{t}(\dot{\Gamma})}\right|^{-1/2}\right)\ .
\label{eqnxi6}
\end{equation}
This function is plotted in Fig.~\ref{quench} for $z=4.5$ and $c=0.83$ (red curves) on a double logarithmic scale together with $\xi_{\textrm{def}}$ and $\xi_{\textrm{dom}}$ at the freeze-out temperature $\hat{\Gamma}$. We find very good agreement. Nonetheless, we fit for comparison the data via an algebraic scaling (blue dotted lines) of the form $f(\dot{\Gamma})=a\dot{\Gamma}^{-\kappa}$, for which we observe $\kappa_{\textrm{def}}=0.061\pm0.001$ and $\kappa_{\textrm{dom}}=0.061\pm0.002$. The data are compatible with the algebraic decay only for intermediate cooling rates. The deviations from standard Kibble-Zurek mechanism for systems with second order transitions are in line with the temperature-quenched 2D $XY$ model~\cite{Jelic2011} having also non algebraic divergences of the correlation length in equilibrium and being thus in a similar universality class. The small algebraic exponent $\kappa$ can be explained by the relatively large value of the dynamical exponent $z$ which regulates the `slope' of $\hat{\xi}_6(\dot{\Gamma})$ (in~\cite{Watanabe2004}, a value $z=2.5$ was proposed for the hard-disk system). This is due to quite long correlation times in this colloidal system (see Fig.~\ref{time}) which are caused by its overdamped dynamics. Note that the `sonic horizon' is set by the sound velocity of the colloidal monolayer (and not the solvent) being  $\sim\,$mm/s, six orders of magnitude slower compared to atomic systems.

\section{Conclusions}

We presented a colloidal model system, where structure formation in spontaneously symmetry broken systems can be investigated with single particle resolution. The theoretical framework is given by the Kibble-Zurek mechanism which describes domain formation on different scales like the Higgs field shortly after the Big Bang or the vortex network in $^4$He quenched into the superfluid state. Along various cooling rates, we analyzed the development of defects and symmetry broken domains when the systems falls out of equilibrium and fluctuations of the order parameter can not follow adiabatically due to critical slowing down. While 2D melting in the colloidal monolayer is described by KTHNY theory where the divergence of the relevant correlation lengths in equilibrium is exponential (rather than algebraic as typically found in 3D systems), the central idea of the Kibble-Zurek mechanism still holds and the scaling of the observed domain network is correctly described. Implicitly this shows that existence of grain boundaries can not solely be used as criterion for first order phase transitions and nucleation or to falsify second/continuous order transitions since they naturally arise for non-zero cooling rates. Those will always be present on finite time scales in experiment and computer simulations after preparation of the system.

\begin{acknowledgments}
P.K. acknowledges fruitful discussion with S$\acute{\textrm{e}}$bastien Balibar and financial support from the German Research Foundation (DFG), KE 1168/8-1.
\end{acknowledgments}

\end{document}